\documentclass{mem}
\usepackage{natbib}\usepackage{txfonts}\usepackage{balance}
\usepackage{graphicx}
\usepackage[a4paper,breaklinks,dvipdfm]{hyperref}
\begin{document}
\def\teff{$T\rm_{eff }$}
\def\kms{$\mathrm {km s}^{-1}$}

\title{
The Infra-Red Telescope on board the THESEUS mission
}

   \subtitle{}

\author{
D. \,G\"{o}tz\inst{1} 
\and 
O. Boulade\inst{1}
\and
B. Cordier\inst{1}
\and
E. Le Floc'h\inst{1}
\and
F. Pinsard\inst{1}
\and
J. Amiaux\inst{1}
\and
T. Tourrette\inst{1}
\and
S. Basa\inst{2}
\and
S. Vergani\inst{3}
\and
J.L. Atteia\inst{4}
\and
S. Covino\inst{5}
\and
G. Ghirlanda\inst{5}
\and
N. Tanvir\inst{6}
\and
A. Blain\inst{6}
\and
P. O' Brien\inst{6}
\and
A. Rossi\inst{7}
\and
G. Stratta\inst{8}
\and
P.G. Casella\inst{9}
\and
E. Bozzo\inst{10}
\and
C. Tenzer\inst{11}
\and
P. Orleanski\inst{12}
\and
L. Amati\inst{7}
          }

\institute{
CEA Saclay - Irfu/D\'epartement d'Astrophysique,
Orme des Merisiers, B\^at. 709,
F-91191 Gif-sur-Yvette, France
\email{diego.gotz@cea.fr}
\and
LAM--Laboratoire d’Astrophysique de Marseille, rue Fr\'ed\'eric Joliot-Curie 38, F-13388 Marseille Cedex 13, France
\and
GEPI, Observatoire de Paris, PSL Research University, CNRS, Place Jules Janssen, F-92190 Meudon, France
\and
IRAP, Université de Toulouse, CNRS, CNES, UPS, (Toulouse), France
\and
INAF, Brera Astronomical Observatory, via Bianchi 46, 23807, Merate (LC), Italy
\and
University of Leicester, Department of Physics \& Astronomy, University Road, Leicester LE1 7RH, UK
\and
INAF, IASF Bologna, via P. Gobetti 101, 40129 Bologna, Italy
\and
Urbino Univeristy, Via S. Chiara 27, 60127 Urbino, Italy
\and
INAF, Osservatorio Astronomico di Roma, Via Frascati 33, 00078 Monteporzio Catone, Italy
\and
ISDC Data Centre for Astrophysics, Chemin d'Ecogia 16, 1290, Versoix, Switzerland
\and
Institut f\"ur Astronomie und Astrophysik, Abteilung Hochenergieastrophysik, Kepler Center for Astro and Particle Physics, Eberhard Karls Universit\"at T\"ubingen, Sand 1, 72076 T\"ubingen, Germany
\and
Space Research Center of the Polish Academy of Sciences, Warsaw, Poland
}

\authorrunning{D. G\"{o}tz et al. }

\titlerunning{The IRT on board THESEUS}

\abstract{
The Infra-Red Telescope (IRT) on board the Transient High Energy Sky and Early Universe Surveyor 
(\textit{THESEUS}) ESA M5 candidate mission will play a key role in identifying and characterizing 
moderate to high redshift Gamma-Ray Bursts afterglows. The IRT is the enabling instrument on board
THESEUS for measuring autonomously the redshift of the several hundreds of GRBs detected 
per year by the Soft X-ray Imager (SXI) and the X- and Gamma-Ray Imaging Spectrometer (XGIS), 
and thus allowing the big ground based telescopes to be triggered
on a redshift pre-selected sample, and finally fulfilling the cosmological goals of the mission.

The IRT will be composed by a primary mirror of 0.7 m of diameter coupled to a single camera
in a Cassegrain design. It will work in the 0.7--1.8 $\mu$m wavelength range, and will 
provide a 10$\times$10 arc min imaging field of view with sub-arc second localization capabilities, and, at the same time, a 5$\times$5 arc min field of view with moderate (R up to $\sim$500) spectroscopic capabilities.  Its sensitivity, mainly limited by the satellite jitter, is adapted to detect all the GRBs, localized by the SXI/XGIS, and to acquire spectra for the majority of them.
}
\maketitle{}

\section{Introduction}

Despite the recent progresses in Gamma-Ray Burst (GRB) science, obtained in particular thanks to the \textit{Swift} and \textit{Fermi} satellites \citep[e.g.]{zhang14}, there are still many open questions in the field. One concerns the mechanisms that power these extreme explosions (in a handful of seconds the isotropic equivalent energy emitted by GRBs spans from 10$^{50}$ to 10$^{54}$ erg, making them the most luminous events in the Universe), which are still unclear after about five decades since their discovery \citep{klebesadel73}. In particular, the content of the relativistic flow that produces the GRBs remains to be investigated -- especially in terms of its bulk Lorentz factor, its magnetization, its baryon loading and their consequences on the possibility of GRBs being the sources of Ultra High Energy Cosmic Rays (UHECRs).

Another open issue concerns the nature of GRBs progenitors: while thanks to a couple of tens of spectroscopic associations with type I b/c Supernovae, long GRBs are currently considered the endpoint of very massive ($>$30--50 M$_{\odot}$) stars evolution (but their potential binary nature is an open issue), the situation is less clear for what concerns the short GRBs \citep{levan16}. The most popular models involve the possibility of a coalescence of two compact objects (black holes or neutron stars), and the first direct proof may be represented by
GRB/GW 170817 for which a gravitational wave signal from two merging neutron stars has been associated with a short flash of gamma ray radiation, although the derived isotropic emitted energy is far below the one measured for previously known short GRBs \citep{fong17}.

In addition to particle acceleration, radiation physics, and stellar evolution, GRBs are also pertinent to other branches of astrophysics, like Cosmology. Indeed, thanks to their bright afterglows they allow the observers to pinpoint the most distant galaxies \citep{perley16}, and to study them thanks to the imprinting on the afterglow data of the GRB close environment and the intervening matter between the GRB and the Earth, gathering a wealth of information on the structure and physical state of the gas in the Universe through spectroscopy.

\begin{figure}[t!]
\resizebox{\hsize}{!}{\includegraphics{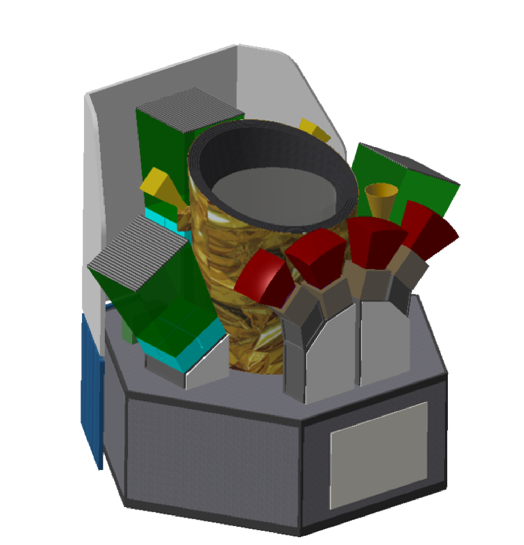}}
\caption{\footnotesize The \textit{THESEUS} satellite. The IRT is at the centre, and the four SXI telescopes are represented in red, while the three XGIS are represented in green.
}
\label{fig:theseus}
\end{figure}

\textit{THESEUS} (Transient High Energy Sky and Early Universe Surveyor) is a project submitted to the European Space Agency (ESA) in the framework of the call
for the fifth medium size mission (M5) of the Cosmic Vision programme. 
One of its main scientific pillars is the study of the early Universe making use of GRB afterglows as beacons \citep{amati17}. 
The selected M5 mission will be flown as early as 2029. The \textit{THESEUS} payload is composed by three instruments: 
a set of three coded mask telescopes, called XGIS for X and Gamma-ray Imaging Spectrometer, a set of four 
wide field X-ray telescopes (Soft X-ray Imager, SXI) making use of ``Lobster Eye'' optics, and an Infra-Red Telescope, the IRT (see Fig. \ref{fig:theseus}). The IRT is the key enabling instrument for the Cosmological goals of the \textit{THESEUS} mission: in fact the SXI (the most sensitive of the two triggering telescopes for distant GRBs) will provide up to 1000 triggers per year, and, thanks to the agility of the \textit{THESEUS} platform (once a trigger is declared on board, a slew request will be issued in order to put the error box in the field of view of the IRT), the IRT will be able to observe them systematically within a few minutes providing a sub-arc second position and a spectroscopic redshift. The same combination of sensitivity, flexibility and availability would not be possible on Earth, unless one would provide a dedicated network of more than ten robotic infra-red telescopes of the 2 m-class, which is a very complex task.

In the following sections the IRT and its performances will be described, as well as the preliminary observation strategy.

\section{The Infra-Red Telescope}

The scientific goals of \textit{THESEUS} (see \citealt{amati17}) require the following on-board capabilities for a near-infrared telescope to follow-up GRB triggers after a demanded spacecraft slew: 

\begin{enumerate}
\item Identify and localize the GRBs found by the SXI and XGIS to arc second accuracy in the visible and near IR domain (0.7--2.0 $\mu$m); 
\item Autonomously determine the photometric redshift of GRBs at redshift z$>$4 and provide redshift upper limits for those at lower redshift; 
\item Provide spectra to precisely measure the redshift, quantify the intrinsic absorption (N$_{H}$), and possibly the metallicity for the majority of GRBs, and in particular those at redshift $>$4; 
\end{enumerate}

The IRT will be provided by a consortium led by France (CEA--Irfu, LAM, IRAP, GEPI, CNES) in collaboration with Switzerland (Universit\'e de Gen\`eve), Germany (IAAT) and Poland (SRC). The IRT will be composed by a primary mirror of 0.7 m of diameter (made of SiC) and a secondary mirror of 0.23 m in a Cassegrain configuration. It's total mass will be of about 110 kg for a power consumption of about 115 W.
The telescope (provided by ESA) will be coupled to a camera (provided by the consortium) based on a 
2048$\times$2048 pixels HgCdTe detector (18 $\mu$m/pixels, resulting in a $\sim$0.3 arcsec/pix plate scale). The detector will be optimised to be sensitive in the 0.7-1.8 (goal 2.0) $\mu$m wavelength range. This choice
is based on the fact that, in order to measure the redshift, z,  of GRB afterglows, the Lyman alpha break 
(at 0.126 $\mu$m at z=0) needs to be clearly identified. The selected wavelength range is hence adapted to measure redshifts between 4 and 10. Nowadays the bulk of GRBs are detected between redshift 2 and 3, hence
this population is already well characterized, while the number of GRBs with measured redshift rapidly decreases above z=4--5. 

In this configuration the expected sensitivity of the IRT will allow for a $\sim$5 $\sigma$ detection of a 20.6 magnitude (H, AB) point like source in a 300 s integration, see Table \ref{tab:IRT}. The telescope sensitivity is partially limited by the platform jitter. Taking the latter into account (1 arc sec jitter over 10 s, at 3 $\sigma$ level), we foresee to limit the image integration time to a maximum of 10 seconds per frame in order to correct for the jitter, and hence such short integration times will induce a high Read-out Noise (RoN) degrading in turn the IRT sensitivity. In addition, due to the absolute pointing error (APE) capability of the platform (2 arc minutes), the high resolution spectroscopy mode cannot make use of a fine slit, and a slit-less mode over a 5$\times$5 arc min area of the detector will be implemented (similarly to what is done for the WFC3 on board the Hubble Space Telescope), with the idea of making use of the rest of the image to locate bright sources in order to correct the frames astrometry a posteriori for the telescope jitter. This is possible thanks to the selectable number of outputs (up to 32) of the Teledyne Hawaii 2RG detector\footnote{Similar performances could be reached by a European 2k$\times$2k detector, currently under development jointly by CEA-LETI and Sofradir and partly funded by ESA, which is expected to reach TRL 6 by 2021.}, which is the current detector baseline. The same goal could also be obtained by making use of the information provided by the high precision star trackers mounted on the IRT. As a consequence the the maximum limiting resolution that can be achieved by such a system for spectroscopy is limited to R$\sim$500 for a sensitivity limit of about 17.5 (H, AB) considering a total integration time of 1800 s. The IRT expected performances are summarized in Table \ref{tab:IRT}, and explained in more detail in Section \ref{sec:irt_perf}, where we show that they are well adapted to the detection of the vast majority of GRB afterglows in the NIR band.

\begin{table*}
\caption{IRT specifications and performances.}
\label{tab:IRT}
\hspace{-1.5cm}
\begin{tabular}{|l|l|}
\hline
Telescope type & Cassegrain \\
\hline
Primary \& Secondary Size & 0.7 m \& 0.23 m\\
\hline
Detector Type & HgCdTe 2048 x 2048 pixels (18  $\mu$m each)\\
\hline
Wavelength range & 0.7-1.8 $\mu$m\\
\hline
Imaging plate scale & 0.3 arc sec/pixel\\
\hline
Field of view & 10$^{\prime}\times$10$^{\prime}$ (imaging mode) | 10$^{\prime}\times$10$^{\prime}$ (low resolution mode) | 5$^{\prime}\times$5$^{\prime}$ (high resolution mode)\\
\hline
Spectral resolution ($ \Delta\lambda$/$ \lambda $) & 2-3 (imaging mode) | 20 (low resolution mode) | 200-500 (high resolution mode)\\
\hline
Sensitivity (AB mag, 5$\sigma$)& H=20.6 (300 s) (imaging) | H=18.5 (300 s) (low resolution) | H=17.5 (1800 s) (high resolution)\\
\hline
Filters & ZYJH (imaging) | Prism (low resolution) | VPH grating (high resolution)\\
\hline
\end{tabular}
\end{table*}

\begin{figure*}[t!]
{\includegraphics[width=0.6   \textwidth]{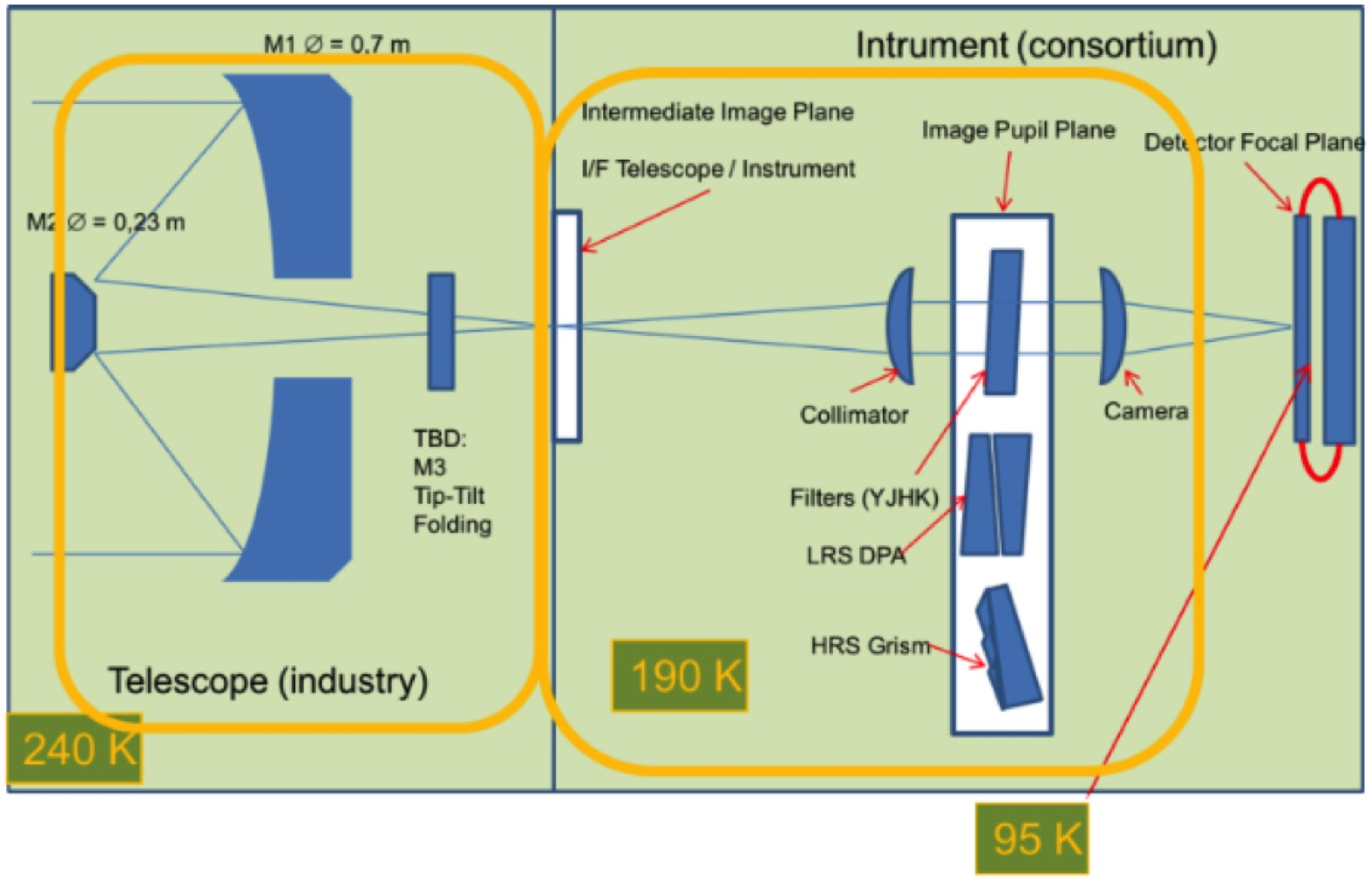}}
{\includegraphics[width=0.4\textwidth]{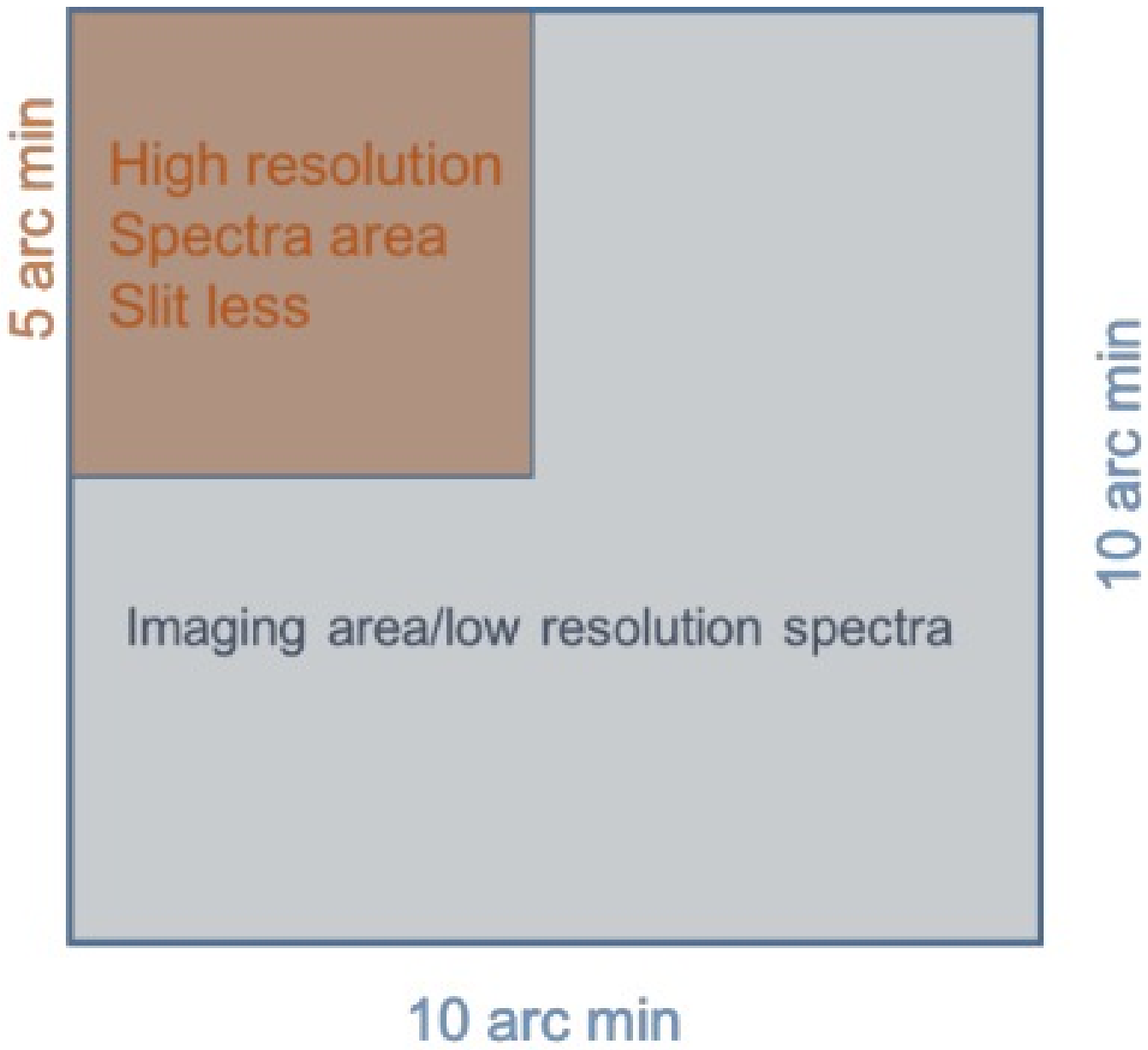}}
\caption{\footnotesize Left: The IRT camera functional scheme. Right: The IRT focal plane: a 10$\times$10 arc min field of view is available for imaging and low resolution spectroscopy, while only a 5$\times$5 arc min field of view is used for slit-less high resolution spectroscopy to limit source confusion.
}
\label{fig:irt_concept}
\end{figure*}

In order to keep the camera design as simple as possible (i.e. avoiding to implement too many mechanisms, like tip-tilting mirrors, moving slits, etc.), we proposed to implement a design with an intermediate focal plane making the interface between the telescope provided by ESA/industry and the IRT instrument provided by the consortium, as shown in the block diagram in Fig. \ref{fig:irt_concept}. The focal plane instrument is composed by a spectral wheel and a filter wheel in which the ZYJH filters, a prism and a volume phase holographic (VPH) grating will be mounted, in order to provide the expected scientific product (imaging, low and high-resolution spectra of GRB afterglows and other transients).

In order to achieve the performances described above (i.e. in conditions such that thermal background represents less than 20\% of sky background) the telescope needs to be cooled at 240 ($\pm$3) K , and this could be achieved by passive means.
Concerning the IRT camera, the optics box needs to be cooled to 190 ($\pm$5) K and the IR detector itself to 95
($\pm$10) K: this allows the detector dark current to be kept at an acceptable level.
Preliminary thermal studies making use of the Earth fluxes computed for the \textit{SVOM}
mission \citep{svom}, which will have a similar orbit in terms of altitude but with a higher inclination (30$^{\circ}$), indicate that use of a Miniature Pulse Tube Cooler (MPTC) coupled to a 0.3 m$^2$ radiator, which periodically faces the Earth atmosphere, would allow to reach the desired temperatures for the camera detector.

\subsection{Observing Sequence}
\begin{figure*}[t!]
{\includegraphics[width=0.5\textwidth]{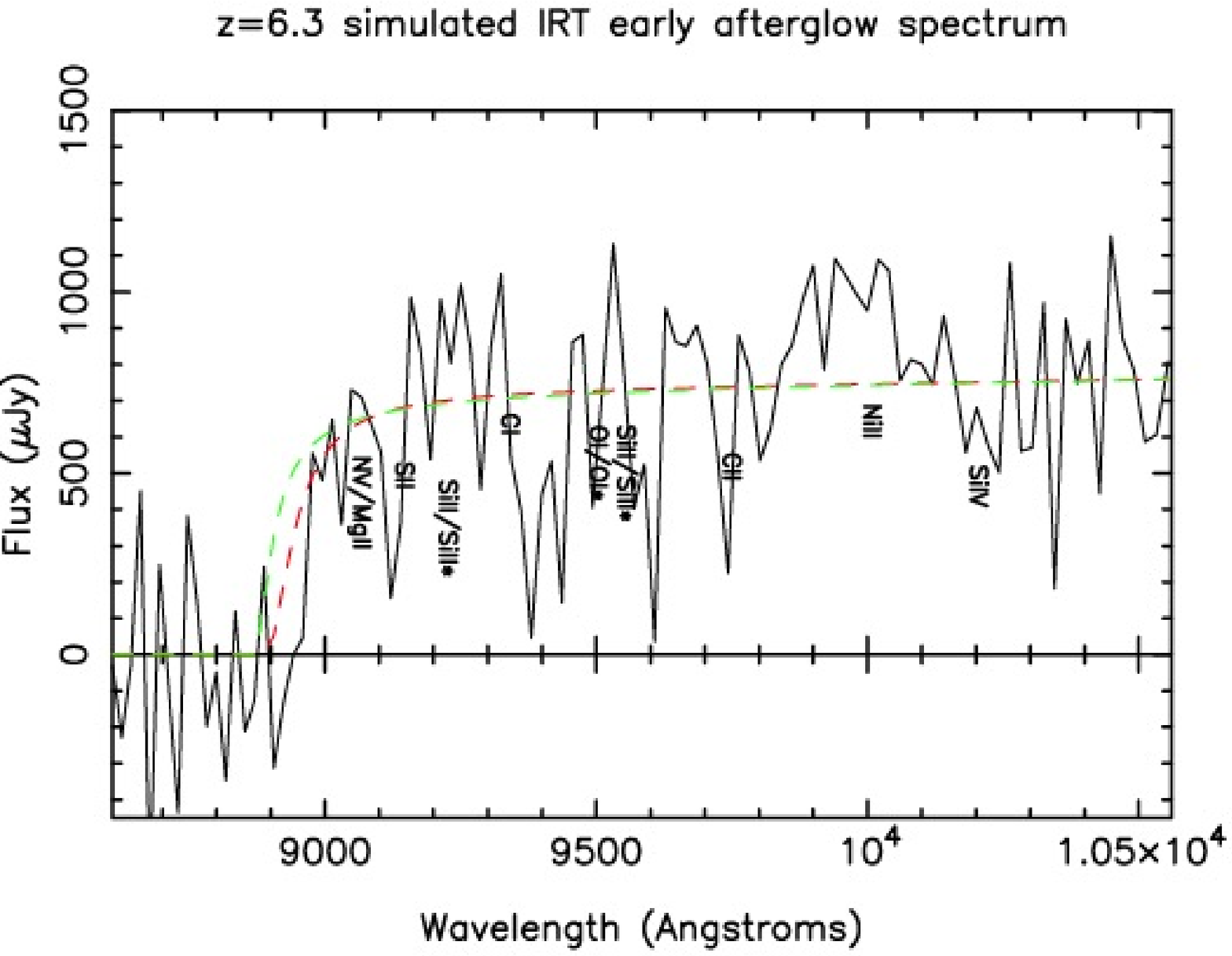}}
\hspace{0.3cm}
{\includegraphics[width=0.5\textwidth]{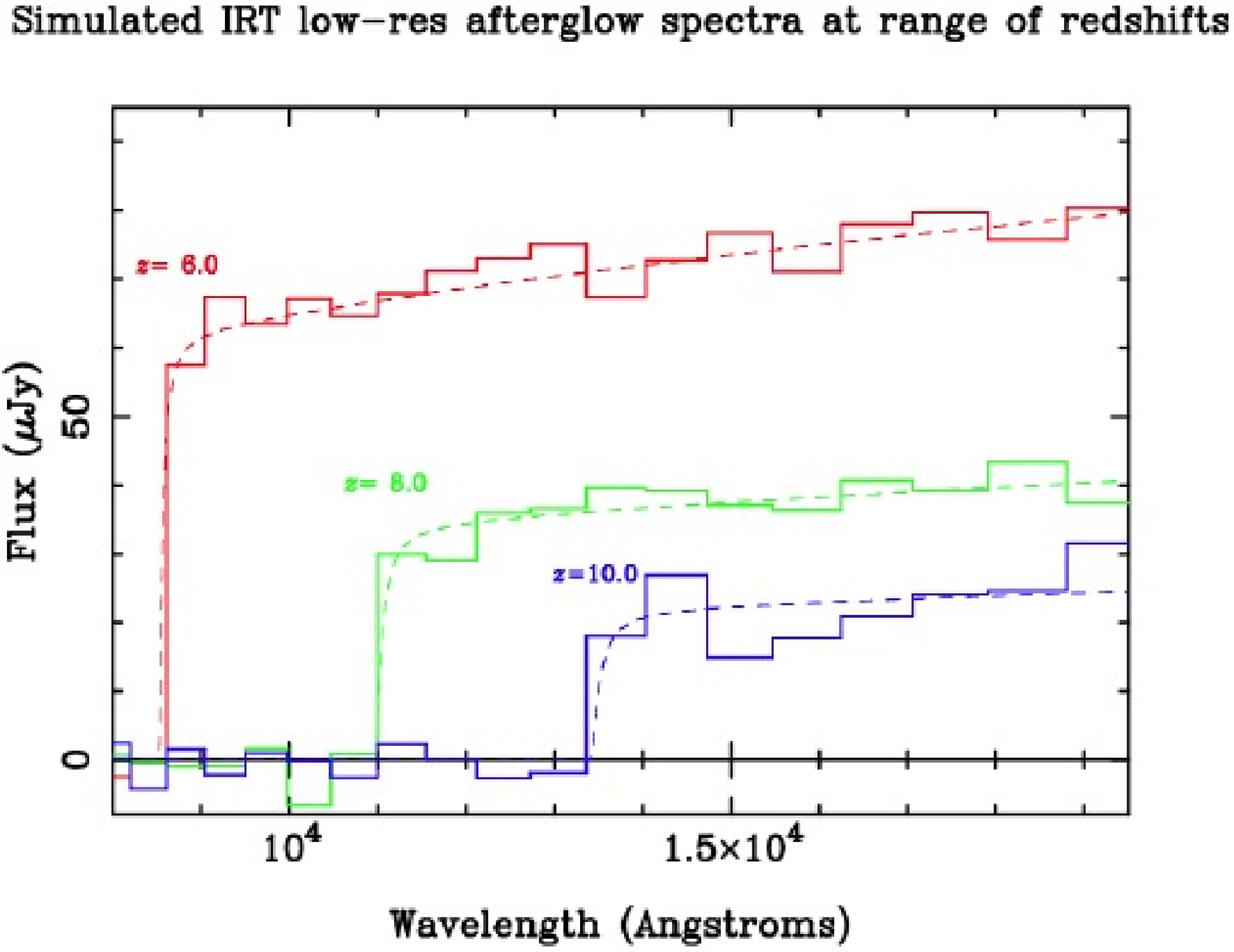}}
\caption{\footnotesize Left: a simulated IRT high resolution (R=500) spectrum for a GRB at z=6.3 observed at 1 hour post trigger assuming a GRB similar to GRB 050904. Right: simulated IRT low resolution (R=20) spectra as a function of redshift for a GRB at the limiting magnitude AB mag 20.8 at z=10, and by assuming a 20 minute exposure. For more details about those simulations, see \citet{amati17}.}
\label{fig:irt_sp}
\end{figure*}

The currently proposed IRT observing sequence for an alert resulting in a platform slew can be summarized as follows:
\begin{enumerate}
\item The IRT will observe the GRB error box in imaging mode as soon as the satellite is stabilized within a few arc sec.
Three initial frames in the ZJH-bands will be taken (10 s each, goal 19 AB 5$\sigma$ sensitivity limit in H) to establish the astrometry and determine the detected sources colours.
\item The IRT will enter the spectroscopy mode (Low Resolution Spectra, LRS) for a total integration time of 5 minutes (expected 5 $\sigma$ sensitivity limit in H 18.5 (AB)), see Fig. \ref{fig:irt_sp}.
\item Sources with peculiar colours and/or variability (such as GRB afterglows) should have been pinpointed while
the low-resolution spectra were obtained and IRT will take a deeper (20 mag sensitivity limit (AB)) H-band image for a total of 60 s. These images will be then added/subtracted on board in order to identify bright variable sources with one of them possibly matching one of the peculiar colour ones. NIR catalogues will also be used in order to exclude known sources from the GRB candidates.
\begin{enumerate}

\item In case a peculiar colour source or/and bright ($<$ 17.5 H (AB)) variable source is found in the imaging step,
the IRT computes its redshift (a numerical value if 4$<$z$<$10 or an upper limit z$<$4--5) from the low resolution spectra obtained at point 1) and determines its position. Both the position and redshift estimate will be sent to ground for follow-up observations. The derived position will then be used in order to ask the satellite to
slew to it so that the source is placed in the high resolution part of the detector plane (see Fig. \ref{fig:irt_concept}) where
the slit-less high resolution mode spectra are acquired. Following the slew, the IRT enters the High Resolution
Spectra (HRS) mode where it shall acquire at least three spectra of the source (for a total exposure time of
1800 s) covering the 0.7-1.8 $\mu$m range, see Fig. \ref{fig:irt_sp}. Then it goes back to imaging mode (H-band) for at least another 1800 s (TBC). 
Note that while acquiring the spectra, continuous imaging is performed on the rest of the detector, see
Fig \ref{fig:irt_concept}. This will allow the on board software to correct the astrometry of the individual frames for satellite drift and jitter, leading to a final correct reconstruction of the spectra by limiting the blurring effects.

\item In case that a faint ($>$ 17.5 H (AB)) variable source is found, IRT computes its redshift from the low resolution
spectra, determines its position and sends both information to the ground (as for 3a). In this case IRT does
not ask for a slew to the platform and stays in imaging mode for a 3600 s time interval to establish the GRB
photometric light curve (covering any possible flaring) and leading the light curve to be known with an
accuracy of $<$5\%.
\end{enumerate}
\end{enumerate}

\section{Scientific Performances}
\label{sec:irt_perf}

\begin{figure}[t!]
\resizebox{\hsize}{!}{\includegraphics{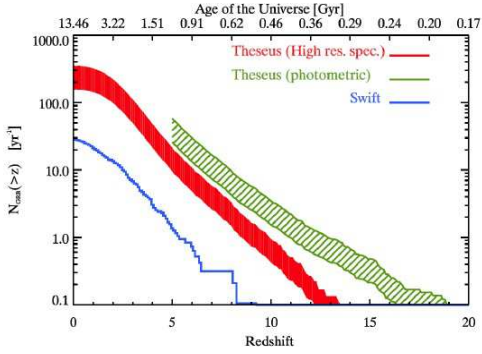}}
\caption{\footnotesize The yearly cumulative distribution of
GRBs with redshift determination as a function of the redshift for \textit{Swift} and \textit{THESEUS}. Details are provided in \citet{amati17}
}
\label{fig:ngrbs}
\end{figure}

The number of expected GRBs triggers per year from the SXI and/or XGIS can reach up to 1000 \citep{amati17}. So one can expect up to three triggers per day. The IRT will follow-up systematically all the triggers providing the associated redshift estimate and possibly high resolution spectra.  This will allow to fulfil the mission requirements, by detecting a large enough number of high-redshift GRBs, see Fig. \ref{fig:ngrbs}. 

IRT will be capable of detecting all afterglows known to date  \citep[see][ for more details]{rossi18} and measure a photometric redshift for $\sim$90\% of the cases (H$>$19.5, AB system), starting observing in LRS mode even 20 min after the trigger. For more distant GRBs (assuming that z$>$6 afterglows have NIR properties similar to those observed up to date) more than 50\% GRB will have H(AB)$<$19.5 5 min after the trigger, allowing for rapid photo-metric and LRS redshift determination. 

There are chances that \textit{THESEUS} could fly at the same time of the next generation X-ray
observatory of ESA, \textit{ATHENA}\footnote{http://www.the-athena-x-ray-observatory.eu}. \textit{ATHENA} will carry a high performance X-ray spectrometer (expected spectral resolution 2.5 eV), X-IFU, that will be used
to study the chemical enrichment of the Universe. Observations of high-z GRB afterglows with the X-IFU 
will allow to simultaneously probe all the elements (C through Ni), in all their ionization stages and different binding states, and thus provides a model-independent survey of the metals during the early phases of the Universe. \textit{THESEUS} will be able to provide redshift selected targets to \textit{ATHENA} (\textit{ATHENA} will implement a rapid response mode, $<$4 hours), and in addition simultaneous NIR spectra obtained with the IRT will provide complementary information about the elemental abundances of the GRB host galaxies.

We note that the only NIR sensitive telescope in space expected for the time of flying of \textit{THESEUS} may be \textit{JWST}\footnote{https://www.jwst.nasa.gov}. While having a great sensitivity, \textit{JWST} will have strong pointing constraints due to thermal reasons, and it is not designed to react quickly on alerts coming from other facilities. On the other hand, \textit{JWST} will be a precious instrument in order to identify and spectroscopically characterize high redshift GRB host galaxies, once the afterglow has faded away.

We also note that all of the next generation ground based OIR facilities -- TMT\footnote{https://www.tmt.org}, GMT\footnote{https://www.gmto.org} and E-ELT\footnote{http://www.eso.org/sci/facilities/eelt/} -- have requirements for rapid response of 30 minutes or less, the most ambitious being the TMT with a 10 minute maximum response time including slewing, instrument and adaptive optics set up, before starting a science exposure. Observatory operational procedures are being
planned that will allow rapid interruption of queue scheduled observations for high priority Target of Opportunity (ToO) programs. Alerts sent to the observatory can be acted upon rapidly via automated processes or on time-scales as short as several minutes by the observing assistants. Some integrated fraction of time, possibly even as high as 10\% to 20\%, will be allotted for ToO programs but the interrupts could occur at any time during queue scheduled observations. However most probably none of these large observatories will be on target faster than the IRT and their schedule will be hardly overruled, unless a firm redshift determination is already available.

\subsection{IRT and Gravitational Waves}

We finally stress that the recent discovery of the electromagnetic counterpart of GW170817 has provided compelling evidence of the existence of kilonovae \citep[e.g.][]{abbott17,pian17}. The temporal and spectral properties of these astrophysical phenomena encode crucial information that goes from the heavy element chemical enrichment of the Universe, the equation of state of neutron stars, the progenitor nature (e.g. NS-NS or NS-BH), etc. The high energy emission preceding a kilonova, e.g. a short GRB and/or an X-ray emission, can be detected with \textit{THESEUS}/XGIS and SXI. However, only with an IRT observation in response to an SXI/XGIS trigger the source sky localization can be refined down arc second level thus enabling large ground-based telescopes as e.g. ELT, to observe and deeply characterize these cosmic events. The observations of GW170817 revealed that at 40 Mpc, the kilonova emission in the J and K bands reached its maximum 1.5 days after the merger with 17.2 mag and 17.5 mag respectively \citep{tanvir17}, thus consolidating the presence of a near infrared emission from such sources, which are well within the IRT capabilities at these distances.

\begin{figure}[t!]
\resizebox{\hsize}{!}{\includegraphics{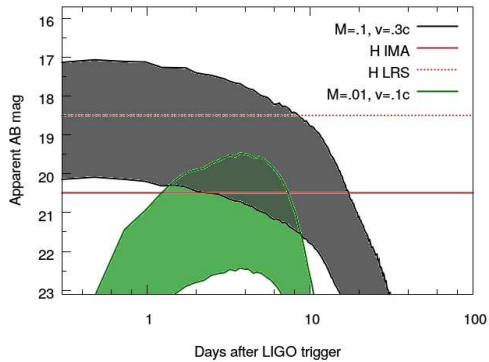}}
\caption{\footnotesize  Predictions of NIR kilonova light curve predictions as a function of distance (the lower curve for 200 Mpc the upper one for 50 Mpc), merger ejecta masses ($M$, in units of solar masses), and velocities ($v$, in units of speed of light) following \citet{barnes16}. The IRT limiting sensitivities for imaging (continuous line) and spectroscopy (dashed line) are shown.
}
\label{fig:kilonova}
\end{figure}

By 2030 the horizon of ground based gravitational wave interferometers will grow, and a comparison between the expected IRT sensitivity and the expected emission from kilonova models between 50 and 200 Mpc is shown in Fig. \ref{fig:kilonova}.
For more details on the \textit{THESEUS} capabilities for multi-messenger astrophysics, see \citet{stratta18}.

\section{IRT Observatory Science}

Based on our conservative estimate of GRB triggers, we expect that 30-40\% of the IRT time will be available and open to the scientific community for additional guest observer (GO) science programs. On one hand, we expect that by the time of the launch of \textit{THESEUS} a wide range of extragalactic science cases will be optimally addressed with the \textit{Euclid}\footnote{http://sci.esa.int/euclid/} survey (e.g. high-z obscured AGN and quasars, high-z clusters). \textit{Euclid} (launch 2021) will cover 15,000 square degrees of the extragalactic sky, down to AB mag(YJH)=24, with R=380 slit-less NIR spectroscopy on the same area; this is $\sim$2 mag deeper than the IRT sensitivity in 1 hr exposure. However, \textit{Euclid} will not have any ability to follow-up transients. In light of this, we will dedicate a fraction of the \textit{THESEUS} observing time to fast Target of Opportunity (ToO) open to the community, in order to broaden our science cases. A spectro-photometric facility to follow-up transient sources provides a unique service to the community at large and is known to yield high-impact science. A hard lesson that we are learning from current surveys is that the number of detected transients is very large and only a minimal fraction of them can be followed-up, resulting in a severe loss of efficiency (as a matter of fact, a large fraction of the newly-discovered transients remains unclassified). In the next decade, time-domain surveys (like CTA, LSST and SKA) and next generation multi-messenger facilities will revolutionise astronomy, providing insight in basically all areas but, at the same time, the need of follow-up facilities will dramatically increase. \textit{THESEUS} will be a key facility to provide the multi-wavelength partner to any kind of transient survey (and will itself work as an X-ray transient factory thanks to the SXI and XGIS instruments). In recent years, at the dawn of time-domain astronomy, the experience of the \textit{Neil Gehrels Swift Observatory} is enlightening. A large fraction of the \textit{Swift} observing time is fully open to ToO observations from the community. Thanks to a flexible planning system, this open time has been optimally exploited, significantly contributing to the success of the \textit{Swift} mission and the advancement in the science of transient and fast variable astrophysical phenomena. However, no other similar space missions with flexible scheduling are currently planned to be in operation around 2030. More importantly, while the astronomical community has (and likely will have in the future) a relatively easy access to optical imaging facilities (mainly ground based) dedicated to the follow-up of transients, the access to facilities operating in the NIR and/or with spectroscopic capabilities will remain difficult. To this end, a multi-frequency space mission equipped with a NIR telescope with spectro-photometric capabilities represents an asset, enabling a fast and efficient precise (sub-arc second) localisation and/or classification of newly-discovered transient sources through multi-frequency SED and spectroscopic characterization with very limited constraints with respect to ground-based observatories (day/night, weather, variable sky background).

A flexible and efficient use of the non-GRB \textit{THESEUS} time with GO and ToO programs will enable to tackle a wide range of studies of variable and transients sources. The near-infrared bands are critical for Solar-system object tracking and multi-epoch variability studies. Cool stars, whose photon fluxes peak in the near-IR, are ideal targets for the detection and characterization of exoplanets using the transit technique, either in surveys or for follow-up
observations of individual sources \citep[e.g.][and references therein]{clanton12}. Simultaneous X-ray and NIR monitoring of samples of T-Tauri stars will shed light on the mechanisms responsible for the onset of the observed outbursts, and how the accretion rate of matter on these stars and the emission of jets can influence the formation of proto-planetary systems. Several open questions for low-mass X-ray binaries, hosting either neutron stars or stellar-mass black holes, require simultaneous IR and X-ray photometry (e.g. concerning the physics of jet emission from these sources; see, e.g., \citealt{migliari10}; \citealt{russell13}). Recent studies have found that the peak luminosity of SNe Ia are genuine standard candles in the NIR \citep[e.g.][]{krisciunas04, burns14}. Considering also the reduced systematics in the NIR related to host-galaxy reddening, the IRT will represent an very efficient tool to construct a low-z sample of SNe Ia to be compared with the high-z samples that will be built by forthcoming IR facilities (e.g. \textit{JWST}).

\section{Conclusions}
The IRT will play a key role in reaching the $\textit{THESEUS}$ scientific goals in the Cosmology field, as well as in the multi-messenger domain. Thanks to the 
implementation of a guest observer programme, it will also represent a powerful tool for the entire astrophysical community, which will benefit of a flexible,
rapid, and sensitive near-infrared telescope in space with both imaging and moderate spectroscopic capabilities.

\begin{acknowledgements}
D.G. acknowledges the financial support of the UnivEarthS Labex program at Sorbonne Paris Cit\'e (ANR-10-LABX-0023 and ANR-11-IDEX-0005-02)
\end{acknowledgements}

\end{document}